\documentclass[12pt]{JHEP3}

\usepackage{cite}
\usepackage{enumerate}
\usepackage{amsbsy}
\usepackage[dvips]{graphicx}
\usepackage{graphics}

\newcommand{\be}{\begin{equation}} \newcommand{\ee}{\end{equation}}
\newcommand{\bea}{\begin{eqnarray}} \newcommand{\eea}{\end{eqnarray}}
\newcommand{\el}{\nonumber \\}
\newcommand{\re}[1]{(\ref{#1})}

\newcommand{\pat}{\partial}

\renewcommand{\sec}[1]{section \ref{#1}}

\newcommand{\brt}[1]{[#1]}
\newcommand{\para}{\paragraph}

\renewcommand{\a}{\alpha}
\renewcommand{\b}{\beta}
\renewcommand{\c}{\gamma}
\renewcommand{\d}{\delta}
\renewcommand{\l}{\lambda}
\newcommand{\mpc}{\mbox{$h^{-1}$Mpc}}

\newcommand{\GN}{G_{\mathrm{N}}}
\newcommand{\ha}{\frac{1}{2}}

\newcommand{\rmd}{\mathrm{d}}
\newcommand{\diag}{\mathrm{diag}}
\newcommand{\bz}{\bar{z}}

\newcommand{\nonum}{\\}
\newcommand {\etal} {et al.}

\newcommand{\adot}{\pat_t{a}}
\newcommand{\addot}{\pat_t^2{a}}
\newcommand{\rhodot}{\dot{\rho}}

\newcommand{\thetadot}{\dot{\theta}}

\newcommand{\bx}{\boldsymbol{x}}

\renewcommand{\H}{\frac{\pat_t a}{a}}
\newcommand{\HH}{\frac{(\pat_t a)^2}{a^2}}

\newcommand{\thetat}{\tilde{\theta}}
\newcommand{\sigmat}{\tilde{\sigma}}
\newcommand{\htt}{\tilde{h}}
\newcommand{\patl}[1]{\frac{\rmd{#1}}{\rmd\l}}
\newcommand{\pate}[1]{\pat_\eta #1}
\newcommand{\path}{\hat{\nabla}}
\newcommand{\ndot}{\dot{n}}
\newcommand{\vdot}{\dot{v}}
\newcommand{\epst}{\tilde\epsilon}
\newcommand{\tdot}{\dot{t}}

\newcommand{\av}[1]{\langle{#1}\rangle}
\newcommand{\sQ}{\mathcal{Q}}
\newcommand{\sR}{{^{(3)}R}}

\newcommand{\sN}{\mathcal{N}}

\newcommand{\PRD}[1]{{\it Phys. Rev.} {\bf D#1}}
\newcommand{\PRE}[1]{{\it Phys. Rev.} {\bf E#1}}
\newcommand{\PRL}[1]{{\it Phys. Rev. Lett.} {\bf #1}}

\newcommand{\NPB}[1]{{\it Nucl. Phys.} {\bf B#1}}
\newcommand{\PLA}[1]{{\it Phys. Lett.} {\bf A#1}}

\newcommand{\MNRAS}[1]{{\it Mon. Not. Roy. Astron. Soc.} {\bf #1}}
\newcommand{\APJ}[1]{{\it Astrophys. J.} {\bf #1}}

\newcommand{\CQG}[1]{{\it Class. Quant. Grav.} {\bf #1}}
\newcommand{\GRG}[1]{{\it Gen. Rel. Grav.} {\bf #1}}
\newcommand{\AaA}[1]{{\it Astron. \& Astrophys.} {\bf #1}}
\newcommand{\PROG}[1]{{\it Prog. Theor. Phys.} {\bf #1}}

\newcommand{\IJMPA}[1]{{\it Int. J. Mod. Phys.} {\bf A#1}}
\newcommand{\IJMPD}[1]{{\it Int. J. Mod. Phys.} {\bf D#1}}
\newcommand{\PRT}[1]{{\it Phys. Rept.} {\bf #1}}

\title{Light propagation in statistically homogeneous and isotropic universes\\ with general matter content}

\author{Syksy R\"{a}s\"{a}nen
\\ Universit\'e de Gen\`eve, D\'epartement de Physique Th\'eorique \\
24 quai Ernest-Ansermet, CH-1211 Gen\`eve 4, Switzerland \\
\email{syksy {\it dot} rasanen {\it at} iki {\it dot} fi}}

\abstract{
We derive the relationship of the redshift and the angular
diameter distance to the average expansion rate for
universes which are statistically homogeneous and isotropic
and where the distribution evolves slowly, but which have
otherwise arbitrary geometry and matter content.
The relevant average expansion rate is selected by
the observable redshift and the assumed symmetry
properties of the spacetime.
We show why light deflection and shear remain small.
We write down the evolution equations for the average expansion
rate and discuss the validity of the dust approximation.
}

\begin{document}
  
\setcounter{tocdepth}{2}

\setcounter{secnumdepth}{3}

\section{Introduction} \label{sec:intro}

\para{A factor of two.}

The early universe from at least big bang nucleosynthesis onwards
is well described by a model where the geometry is locally
spatially homogeneous and isotropic up to linear perturbations,
the matter consists of a gas of particles with
positive pressure, and the relation between geometry
and matter is given by the Einstein equation
based on the four-dimensional Einstein-Hilbert action.
However, at late times such a model underpredicts the
distance to far-away sources and the expansion rate.
Compared to the simplest possibility, the spatially flat
matter-dominated model, the discrepancy is a factor
of about two in both the distance
(for a fixed Hubble constant) and the expansion rate
(for a fixed energy density or age of the universe).
Therefore at least one of the three assumptions
--homogeneity and isotropy, standard matter content
and standard gravity-- is wrong, assuming that light
propagation is correctly modeled by null geodesics.
No deviations from standard gravity have been observed in
local physics, not in the solar system
(apart from the Pioneer anomaly and the flyby anomaly, where
the possibility of systematics is not ruled out) nor
in pulsars \cite{solar, Will:2005}.
Neither is there any detection of an effect of exotic
matter with negative pressure on local physics.
The factor of two discrepancy only appears in
observations of distance and expansion rate
which involve quantities integrated over
large scales\footnote{It has been argued that locally
repulsive gravity has been observed in the motions
of galaxies near the Local Group \cite{localde}.
This is an interesting possibility, but the present data
is not precise enough for such a detection.}.
This situation is quite different from that of dark matter,
for which there is evidence from various different systems
on several scales.

While there is no evidence against standard general relativity
or standard matter apart from the increased distance and
expansion rate,
the universe is known to be locally far from homogeneity and
isotropy due to the formation of non-linear structures at late
times. It is possible that the breakdown of the homogeneous
and isotropic approximation could explain the failure of
the prediction of homogeneous and isotropic models with
ordinary matter and gravity
\cite{Buchert:2000, Tatekawa:2001, Wetterich:2001, Schwarz, Rasanen, Kolb:2004}.
The effect of inhomogeneity and/or anisotropy on the
evolution of the universe was first discussed in detail
in \cite{fitting} under the name ``the fitting problem'',
and the effect on the average expansion rate is known as
backreaction
\cite{Buchert:1999, Buchert:2001, Ellis:2005, Rasanen:2006b, Buchert:2007}.
It has been shown with toy models that inhomogeneities
can lead to accelerating expansion
\cite{Rasanen:2006b, Chuang:2005, Paranjape:2006a, Kai:2006, Rasanen:2006a},
but whether this happens for the distribution of structures
present in the real universe is not yet clear.
The order of magnitude of the observed change in
the expansion rate and the correct timescale
of around 10 billion years do emerge from the physics of
structure formation in a semi-realistic model
\cite{Rasanen:2008a, peakrevs}, but there is no fully
realistic calculation yet.

\para{Light propagation and statistical homogeneity and isotropy.}

Most cosmological observations probe quantities
related to light propagation, such as the redshift,
the angular diameter distance (or equivalently the
luminosity distance) and image distortion.
In linearly perturbed homogeneous and isotropic
Friedmann-Robertson-Walker (FRW) models, the redshift
and the distance are to leading order determined by the
expansion rate and the spatial curvature.
The corrections due to the perturbations are small for
typical light rays, and are important only
for the cosmic microwave background (CMB), whose redshift
anisotropies are very accurately measured, and image
distortion, which is zero for the background and remains
small when perturbations are included. (At least this
is the case when the perturbations are statistically
homogeneous and isotropic and the homogeneity scale is small;
see \cite{Enqvist:2009} for a counterexample with a large
spherically symmetric structure.)

The fact that the optical properties of a FRW universe can be
expressed in terms of the expansion rate and the spatial curvature
is rather obvious, because light propagation is, for small
wavelengths, purely geometrical, and these are the only
degrees of freedom in the FRW geometry.
In a general spacetime, the situation is more involved.
Nevertheless, if the distribution of the geometry is statistically
homogeneous and isotropic, it could be expected that light
propagation over distances longer than the homogeneity scale
can to a first approximation be similarly described with a few
quantities related to the overall geometry, regardless of complicated
local details \cite{Rasanen:2008a}.
Light propagation in statistically homogeneous and isotropic
universes with irrotational dust as the matter content was studied
in \cite{Rasanen:2008b}, where it was argued that if the distribution
evolves slowly compared to the time it takes for light
to cross the homogeneity scale, then the redshift and the
angular diameter distance are determined by the
expansion rate as a function of redshift and the
matter density today.
The study \cite{Rasanen:2008b} had three shortcomings.

First, it was assumed that the variation in the spatial
direction of the null geodesics (i.e. light deflection) is small.
The magnitude of the null shear was also left undetermined.
Observationally, both light deflection and image distortion
are known to be small for typical light rays \cite{Munshi:2006},
and it should be established that this follows from the
symmetry properties of the spacetime.
Second, the treatment of matter as irrotational dust
is not locally valid \cite{Buchert:2005, Pueblas:2008},
because effects such as rotation and velocity dispersion are
important for stabilising structures on small scales.
The vorticity and non-dust nature of the matter content
may be expected to be unimportant for the overall cosmological
evolution in the real universe at late times.
Nevertheless, such effects should be included to establish under
which conditions they can be neglected, and put the dust
approximation on better footing. Including matter
other than dust is also necessary for treating backreaction
in the early universe such as during inflation
\cite{Buchert:2001, Woodard, Unruh, Geshnizjani, Brandenberger:2002}
or preheating.
Third, the arguments were qualitative and the corrections
to the mean behaviour were not determined.

We now remedy the first and second problems.
We derive results for light propagation, including
deflection and shear, using only assumptions
about the symmetry of the spacetime geometry and matter content.
We consider general matter content and include rotation.
Concentrating on observable quantities related to light propagation,
we show how the relevant averaging hypersurface is
given by the statistical symmetry of the spacetime.
However, our analysis is not more quantitative
than \cite{Rasanen:2008b}, and the arguments should
be followed up with a more rigorous study.

In section 2 we go through our assumptions, set up
the covariant formalism and derive
results for the redshift, the deflection,
the null shear and the angular diameter distance.
In section 3 we derive the evolution equations
for the scale factor, which generalise the
Buchert equations of the irrotational
dust case \cite{Buchert:1999}, and consider
the validity of the dust approximation.
In section 4 we discuss the possible effect
of the discreteness of the matter content,
the relevance of average quantities and the FRW
description, and summarise the situation.

\section{Light propagation} \label{sec:light}

\subsection{Spacetime geometry} \label{sec:geom}

\para{Statistical symmetry.}

We assume that there exists a foliation of the
spacetime into spatial hypersurfaces of statistical
homogeneity and isotropy, which we denote by $\sN$.
The time which is constant on such a hypersurface
is denoted $t$, and when referring to a particular
hypersurface, we use the notation $\sN(t)$.
By this we mean that when we consider
any region larger than the homogeneity scale, the average
quantities within the region do not depend on its
location, orientation or size.
In other words, over large scales, there are no
preferred locations or directions, and no correlations.
Locally the dynamics can be complex, as the assumption
of statistical homogeneity and isotropy only concerns
average quantities evaluated over large scales.
The frame of statistical homogeneity and isotropy
may not locally coincide with either
the Eckart frame (where there particle number flux is zero) or
the Landau-Lifshitz frame (where the energy flux is zero)
\cite{Maartens:1998}.
However, statistical homogeneity and isotropy does imply
that the integrated flux of any quantity through the
boundary of a volume larger than the homogeneity scale
vanishes.

In this view, the universe consists of identical (up
to statistical fluctuations) boxes stacked next
to each other.
In the real universe, there are correlations even
over scales longer than the Hubble scale,
due to inflation (or some other process in the early universe)
which produces a large region that is exactly homogeneous
and isotropic except for linear perturbations.
The distribution of the perturbations
is statistically homogeneous and isotropic.
When the perturbations become non-linear at late
times (in typical supersymmetric dark matter models,
the first structures form around a redshift of 40--60 \cite{SUSYCDM}),
local homogeneity and isotropy are lost, but the distribution
of non-linear structures remains statistically homogeneous
and isotropic, and the amplitude of
correlations is small beyond the homogeneity scale.
What one finds as the homogeneity scale depends on
the limit that one sets for this amplitude.
Based on the fractal dimension of the point set of galaxies,
it has been argued that the distribution becomes homogeneous on a
scale of around 100 Mpc to an accuracy of about 10\% \cite{hom}.
However, there are still large fluctuations on 100 Mpc scales,
and it has been argued that the sample size is not large
enough to establish that the distribution is self-averaging,
which is a necessary condition for statistical homogeneity
\cite{inhom, SylosLabini:2009}.
Studies of morphology also suggest that the homogeneity
scale could be 300 Mpc or more \cite{morphology}.

Statistical homogeneity and isotropy is
formulated in terms of spatial hypersurfaces, but
light travels along null geodesics, not in a spacelike
direction. Therefore we also need information about the
evolution which relates one hypersurface to the next.
We assume that the evolution is slow in the sense
that the timescale of change in the spatial distribution
is much larger than the homogeneity scale.
Phrased differently, the variation of the geometry
along a null geodesic is rapid compared to the
scale over which the mean varies significantly.
In the real universe, the timescale for change
in the distribution of matter and geometry
is the Hubble time $H^{-1}$.
Today $H_0^{-1}=3000$\mpc{} (with $h$ somewhat below unity
\cite{Hubble}), much larger than 100--300 Mpc.
In the past, the homogeneity scale was even smaller
relative to the Hubble scale, as structure formation
was less advanced.

The combination of statistical homogeneity and
isotropy on spatial hypersurfaces and slow evolution
from one hypersurface to the next can be heuristically
thought of as a distribution that is statistically
approximately homogeneous and isotropic in four dimensions
when considering scales larger than the homogeneity scale,
but smaller than the timescale of change in the distribution.
The notion of statistical homogeneity and isotropy
in general spacetimes should be made more rigorous,
and the role of slow evolution in the arguments we
make below on light propagation should be quantified.

\para{The two frames.}

We denote the vector normal to $\sN$ by $n^\a$ and
the velocity of the observers by $u^\a$.
Both are normalised to unity, $n_\a n^\a=u_\a u^\a=-1$.
The observer velocity is completely general, it
is not assumed to be geodesic or irrotational.
For reviews of the covariant approach we use, see
\cite{Ehlers:1961, Ellis:1971, Ellis:1998c, Clarkson:2000, Tsagas:2007};
for the relation to the ADM formalism \cite{Arnowitt:1962},
see \cite{Jantzen:2001}.
The tensors which project on the hypersurface
orthogonal to $n^\a$ and the rest space
orthogonal to $u^\a$ are, respectively,
\bea \label{h}
  h_{\a\b} &\equiv& g_{\a\b} + n_\a n_\b \el
  h_{\a\b}^{(u)} &\equiv& g_{\a\b} + u_\a u_\b \ ,
\eea

\noindent where $g_{\a\b}$ is the spacetime metric.
The restriction of the projection tensor
$h_{\a\b}$ to $\sN$ is the metric on $\sN$.
The spatial derivative of a scalar is defined as
$\path_\a f\equiv h_{\a}^{\ \b} \nabla_\b f$, for vectors we have
$\path_\b f_\a \equiv h_{\b}^{\ \d} h_{\a}^{\ \c} \nabla_\d f_\c$,
and similarly for higher order tensors.
The spatially projected traceless part of a tensor is
$f_{\langle\a\b\rangle}\equiv h_{(\a}^{\ \ \c} h_{\b)}^{\ \ \d} f_{\c\d} - \frac{1}{3} h_{\a\b} h^{\c\d} f_{\c\d}$.
The volume element on $\sN$ is $\epsilon_{\a\b\c}\equiv\eta_{\a\b\c\d} n^\d$, 
where $\eta_{\a\b\c\d}$ is the spacetime volume element.
The derivative with respect to the proper time $s$
of the frame of statistical homogeneity and isotropy
is $n^\a\nabla_\a$, and it is denoted by an overdot.
Unless $\ndot^\a=0$, the proper time $s$ does not coincide
with the time $t$ which is constant on $\sN$.
We can write $n_\a=-\tdot^{-1}\pat_\a t$.
The derivative with respect to $t$ is
$m^\a \nabla_\a$, with $m^\a=\tdot^{-1} n^\a$.
We define $\Gamma\equiv-n_\a m^\a$, so
$\Gamma=\tdot^{-1}=\pat_t s$ and $m^\a=\Gamma n^\a$.
Physically, $\Gamma$ describes the time dilation due to
the non-geodesic motion of the $n^\a$ frame; we have
$\ndot_\a=\path_\a\ln\Gamma$.
(Note that a $\Gamma$ which depends only on $t$
corresponds to a different time coordinate, not
different physics.)
Only if $\ndot^\a=0$ can we choose $\Gamma=1$ and $s=t$,
which is equivalent to the statement that $\sN$ is a
hypersurface of constant proper time.
In addition to $s$ and $t$, we also have the
proper time of the observers, defined by $u^\a$.

Because the timescale for the evolution of structures
is determined by their proper time, the hypersurface of statistical
homogeneity and isotropy could be expected to coincide with the
hypersurface of constant proper time of observers
comoving with the structures, as argued in
\cite{Rasanen:2006b, Rasanen:2008a, Rasanen:2008b}.
However, if the matter consists of several components
which form structures differently, the situation is not
so simple. For example, in the real universe,
dark matter and baryons cluster differently (though the differences
are not expected to be important on scales larger
than the homogeneity scale).
And on small scales, dark matter is multistreaming,
so there is more than one proper time
associated with the matter flow at a single point.
We keep the hypersurface of statistical homogeneity
and isotropy arbitrary.

Without loss of generality, we write the observer
velocity $u^\a$ in terms of $n^\a$ and a component
orthogonal to $n^\a$,
\bea \label{n}
  u^\a = \gamma ( n^\a + v^\a ) \ ,
\eea

\noindent where $\gamma\equiv -n_\a u^\a=(1-v^2)^{-1/2}$, with
$v^2\equiv v_\a v^\a$ and $v_\a n^\a=0$.
Note that $v^\a$ is not the peculiar velocity, either
in the perturbation theory sense of a velocity with
respect to a fictitious background, or in the
physical sense of deviation from
a shearfree velocity field \cite{peculiar}.
The quantity $v^\a$ measures the deviation of the local
observer velocity from the time direction set by the
frame of statistical homogeneity and isotropy.
Even if $v^\a$ is zero, there can be arbitrarily large
spatial variations in the expansion rate.
We will see that a large $v$ implies significant anisotropy
in the CMB. We therefore often take $v$ to be small,
and expand to first order in $v$. We use $\simeq$ to indicate
equality up to and including terms first order in $v^\a$.
(We do not assume that derivatives of $v^\a$ are small.)
Physically, this means that the motion of the observers
with respect to the frame of homogeneity and isotropy is
non-relativistic.

\para{Fluid kinematics.}

The covariant derivative of $n^\a$ can be decomposed as
\bea \label{gradn}
  \nabla_\b n_\a
  &=& \frac{1}{3} h_{\a\b} \theta + \sigma_{\a\b} - \ndot_\a n_\b \ ,
\eea

\noindent where
$\theta\equiv\nabla_\a n^\a=\path_\a n^\a$
is the volume expansion rate and
$\sigma_{\a\b}\equiv\nabla_{\langle\b} n_{\a\rangle}=\path_{\langle\b} n_{\a\rangle}$
is the shear tensor.
The tensor $\sigma_{\a\b}$ and the acceleration vector
$\ndot^\a$ are spatial in the sense that they are orthogonal
to $n^\a$, $\sigma_{\a\b} n^\b=0$, $\ndot_\a n^\a=0$.
The shear scalar is defined as
$\sigma^2\equiv\ha\sigma_{\a\b}\sigma^{\a\b}$.
Because $n^\a$ is hypersurface-orthogonal, it follows
from Frobenius' theorem that the vorticity 
$\omega_{\a\b} \equiv \nabla_{[\b} n_{\a]} + \ndot_{[\a} n_{\b]}=\path_{[\b} n_{\a]}$
is zero
\cite{Ehlers:1961, Ellis:1971, Ellis:1998c}, \cite{Wald:1984} (page 434).

The covariant derivative of the observer velocity $u^\a$ can be
analogously decomposed with respect to itself,
\bea \label{gradu}
  \nabla_\b u_\a
  &=& \frac{1}{3} h^{(u)}_{\a\b} \theta^{(u)} + \sigma^{(u)}_{\a\b} + \omega^{(u)}_{\a\b} - A_\a u_\b \ ,
\eea

\noindent where $\theta^{(u)}\equiv\nabla_\a u^\a$,
$\sigma_{\a\b}^{(u)}\equiv h^{(u)}_{\a\c} h^{(u)}_{\b\d} \nabla^\d u^\c-\frac{1}{3}\theta^{(u)} h^{(u)}_{\a\b}$,
$\omega^{(u)}_{\a\b} \equiv \nabla_{[\b} u_{\a]} + A_{[\a}u_{\b]}$
and $A^\a\equiv u^\b\nabla_\b u^\a$.

Given \re{n}, the expansion rates in the two frames 
are related as (see \cite{Tsagas:2007} for the expressions
for the acceleration, shear and vorticity)
\bea
  \label{reltheta} \theta^{(u)} &=& \gamma \theta + \gamma ( \path_\a v^\a + \ndot_\a v^\a ) + \gamma^3 ( \vdot_\a v^\a + v^\a v^\b \path_\a v_\b ) \el
  &\simeq& \theta + \path_\a v^\a + \ndot_\a v^\a + \vdot_\a v^\a \ .
\eea

\para{The energy-momentum tensor.}

In the geometrical optics approximation, light propagation
is kinematical, and independent of the laws which determine
the evolution of the geometry.
However, we prefer to replace the Einstein tensor
with the energy-momentum tensor which describes the
matter content, and to do that we assume that the geometry
is related to the matter by the Einstein equation,
\bea \label{Einstein}
  G_{\a\b} &=& 8 \pi\GN T_{\a\b} \ ,
\eea

\noindent where $G_{\a\b}$ is the Einstein tensor, $\GN$
is Newton's constant, and $T_{\a\b}$ is the energy-momentum tensor.

Without loss of generality, the energy-momentum tensor can
be decomposed with respect to $n^\a$ as
\bea \label{emdecn}
  T_{\a\b} = \rho^{(n)} n_\a n_\b + p^{(n)} h_{\a\b} + 2 q^{(n)}_{(\a} n_{\b)} + \pi^{(n)}_{\a\b} \ ,
\eea

\noindent where $\rho^{(n)}\equiv n^\a n^\b T_{\a\b}$ is the energy density,
$p^{(n)}\equiv\frac{1}{3} h^{\a\b} T_{\a\b}$ is the pressure,
$q^{(n)}_\a\equiv -h_\a^{\ \b} n^\c T_{\b\c}$ is the energy
flux and
$\pi^{(n)}_{\a\b}\equiv h_{\a}^{\ \c} h_{\b}^{\ \d} T_{\c\d} - \frac{1}{3} h_{\a\b} h^{\c\d} T_{\c\d}=T_{\langle\a\b\rangle}$
is the anisotropic stress.
Both $q^{(n)}_\a$ and $\pi^{(n)}_{\a\b}$ are spatial in
the sense that $q^{(n)}_\a n^\a=0, \pi^{(n)}_{\a\b} n^\b=0$.
The quantities measured by the observers
are given by the decomposition with respect to $u^\a$,
\bea \label{emdecu}
  T_{\a\b} = \rho^{(u)} u_\a u_\b + p^{(u)} h^{(u)}_{\a\b} + 2 q^{(u)}_{(\a} u_{\b)} + \pi^{(u)}_{\a\b} \ ,
\eea

\noindent where $\rho^{(u)}, p^{(u)}, q^{(u)}_\a$ and $\pi^{(u)}_{\a\b}$
are defined analogously to the $n^\a$ frame quantities.
Locally, dust is defined as matter for which
$p^{(u)}$, $q^{(u)}_\a$ and $\pi^{(u)}_{\a\b}$ are zero;
it then follows from the equations of motion that $A^\a$
is also zero.
In the $u^\a$ frame, the non-dust terms have a clear physical
interpretation in terms of what the observers measure.
Such terms can arise from the properties of matter
(it may be that the matter cannot be treated as dust in any frame)
and from the fact that an ideal fluid looks non-ideal to a
non-comoving observer. We discuss treating the matter
approximately as dust in \sec{sec:evo}.

We could equally take \re{emdecn} and \re{emdecu}
as decompositions of the Einstein tensor rather than the
energy-momentum tensor.
We use assumed symmetry properties of \re{emdecn}
such as the absence of preferred directions over large
distances, and these could be equally phrased in terms of the
geometry expressed in $G_{\a\b}$.
However, $T_{\a\b}$ is more transparent because it can be
understood in terms of a matter model.

\subsection{Photon energy and redshift}

\para{The photon momentum.}

We want to relate quantities integrated along null
geodesics to average quantities which characterise
the spatial geometry.
We use assumptions about the symmetry properties
of the spacetime, so averages are most meaningfully
discussed in terms of quantities on $\sN$
and the vector $n^\a$.
In contrast, the observable redshift and light
deflection are defined by the observer velocity $u^\a$.
(The angular diameter distance and the null shear
scalar are independent of the velocity field \cite{Sachs:1961}.)

In the geometrical optics approximation light travels on null geodesics
\cite{Misner:1973} (page 570), \cite{Schneider:1992} (page 93).
We do not consider caustics, which are not expected to be
important for typical light rays in cosmology (though see \cite{caustic}).
For treatment of the CMB in the covariant formalism, see
\cite{Maartens:1998, Dunsby:1997, Zibin:2008a}.
The null geodesic tangent vector
is given by the gradient of the phase of the wave,
identified with the photon momentum, and denoted by $k^\a$.
It satisfies $k_\a k^\a=0$ and $k^\a \nabla_\a k^\b=0$.
The redshift plus one is proportional to the energy
measured by the observer, $1+z\propto E^{(u)}$, which in turn is
\bea \label{Eu}
  E^{(u)} = - u_\a k^\a \ .
\eea

The photon momentum can be decomposed into an
amplitude and the direction, and the direction
can be split into components parallel and orthogonal to $u^\a$,
\bea \label{kdecu}
  k^\a = E^{(u)} ( u^\a + r^\a ) \ ,
\eea

\noindent with $u_\a r^\a=0$, $r_\a r^\a=1$.

Because the vector $n^\a$ is adapted to the
symmetry of the spacetime, it is more convenient
to calculate quantities in the $n^\a$ frame and then transform
to the $u^\a$ frame.
The decomposition of $k^\a$ with respect to $n^\a$ reads
\bea \label{kdecn}
  k^\a = E^{(n)} ( n^\a + e^\a ) \ ,
\eea

\noindent with $E^{(n)} \equiv - n_\a k^\a, n_\a e^\a=0$, $e_\a e^\a=1$.
The quantities $E^{(n)}$ and $e^\a$ do not have a
straightforward observational interpretation, unlike $E^{(u)}$ and $r^\a$.

The observed energy $E^{(u)}$ is given in terms of $E^{(n)}$ by
\bea \label{EuEn}
  E^{(u)} &=& \gamma ( 1 - v_\a e^\a ) E^{(n)} \ ,
\eea

\noindent and the observed direction $r^\a$ is related to $e^\a$ by
\bea \label{re}
  r^\a &=& \frac{1}{\gamma ( 1 - v_\b e^\b )} ( n^\a + e^\a ) - \gamma ( n^\a + v^\a ) \el
  &\simeq& ( 1 + v_\b e^\b ) e^\a + v_\b e^\b n^\a - v^\a \ .
\eea

\noindent The inverse relation is
\bea \label{er}
  e^\a &=& \frac{1}{ \gamma + v_\b r^\b } ( u^\a + r^\a ) - \gamma^{-1} u^\a + v^\a \el
  &\simeq& ( 1 - v_\b r^\b ) r^\a - v_\b r^\b u^\a + v^\a \ .
\eea

\para{Statistical homogeneity and isotropy.}

We obtain the evolution of $E^{(n)}$ by operating with the derivative
along the null geodesic, $\patl{}\equiv k^\a\nabla_\a$.
Denoting $\pate{} \equiv (n^\a + e^\a) \pat_\a$ and using
\re{gradn} and \re{kdecn}, we have
\bea \label{Eder}
  E^{(n)} \pate{E^{(n)}} &=& k^\b \nabla_\b E^{(n)} \el
  &=& - k^\a k^\b \nabla_\b n_\a \el
  &=& - {E^{(n)}}^2 \left( \frac{1}{3} \theta + \ndot_\a e^\a + \sigma_{\a\b} e^\a e^\b \right) \ ,
\eea

\noindent which integrates into
\bea \label{Eint}
  E^{(n)}(\eta) &=& E^{(n)}(\eta_0) \exp\left( \int_{\eta}^{\eta_0} \rmd \eta \left[ \frac{1}{3} \theta + \ndot_\a e^\a + \sigma_{\a\b} e^\a e^\b \right] \right) \el
  &=& E^{(n)}(t_0,\bx_0) \exp\left( \int_{t}^{t_0} \rmd t \Gamma \left[ \frac{1}{3} \theta + \ndot_\a e^\a + \sigma_{\a\b} e^\a e^\b \right] \right) \el
  &\approx& E^{(n)}(t_0,\bx_0) \exp\left( \int_{t}^{t_0}\rmd t \frac{1}{3} \av{ \Gamma \theta } \right) \ ,
\eea

\noindent where the integral is along the null geodesic and
the subscript $0$ refers to the observer's position and time.
On the second line we have taken the time $t$ as the integration
variable; the spatial coordinates $\bx$ on $\sN$ are understood
as functions of $t$ on the null geodesic.
We have then taken into account
that if there are no preferred directions in the
geometry of $\sN$ over long distances, and the
direction $e^\a$ changes only little or
evolves much more slowly than the distribution
of the geometry (we discuss this in \sec{sec:def}),
the dominant contribution is given by the average
expansion rate. (We use the symbol $\approx$
to indicate dropping terms which are suppressed due
to statistical homogeneity and isotropy, in contrast
to $\simeq$, which indicates dropping terms which
are small because $v\ll1$.)
The argument for this is the following \cite{Rasanen:2008b}.
If $\Gamma\ndot^\a$ has no preferred orientation,
it points equally in the directions along
and opposite to $e^\a$, so its contribution vanishes.
Similarly, $\Gamma\sigma_{\a\b}$ contributes only via
its trace, which is zero.
The term $\frac{1}{3}\Gamma\theta$ then gives the dominant contribution.
Under the assumption that the timescale for the evolution of
the distribution of the geometry is much larger than the time
it takes for light to cross the homogeneity scale, the
integral is dominated by the average value of $\Gamma\theta$,
as the contributions of the variation around the average cancel.
(See \sec{sec:av} for details of the averaging.)
In reality, there is some evolution of the quantities
along the null geodesic, and the cancellations are not
perfect, so the contributions of $\Gamma\ndot_\a e^\a$,
$\Gamma\sigma_{\a\b} e^\a e^\b$ and of the variation of
$\Gamma\theta$ are only suppressed instead of zero.

The observed energy $E^{(u)}$ is, using \re{EuEn},
\bea
  E^{(u)} &\approx& E^{(n)}(t_0,\bx_0) \gamma ( 1 - v_\a e^\a ) \exp\left( \int_{t}^{t_0}\rmd t \frac{1}{3} \av{ \Gamma \theta } \right) \ ,
\eea

\noindent and the redshift $1+z=E^{(u)}(\eta)/E^{(u)}(\eta_0)$ is
\bea \label{z}
  1+z &\approx& \frac{ \gamma ( 1 - v_\a e^\a ) }{ \gamma_0 ( 1 - v_\a e^\a )|_0 } \exp\left( \int_{t}^{t_0}\rmd t \frac{1}{3} \av{ \Gamma \theta } \right) \ .
\eea

\noindent Expressing $e^\a$ in terms of the observed direction
$r^\a$ with \re{er}, it is transparent that there are
large observed anisotropies in the redshift of isotropically
distributed sources unless $v$ is small or constant or there
is a conspiracy of cancellations.
Conversely, if $v$ is small, the anisotropy is small,
even though the variations in the geometry can be large.
In particular, the near-isotropy of the CMB does
not imply that the universe would be nearly FRW
\cite{Rasanen:2008b, Rasanen:2009}.
Assuming $v\ll1$, the correction due to $v$ reduces to 
$v_\a e^\a|_0-v_\a e^\a\simeq v_\a r^\a|_0-v_\a r^\a$.
The first term is the dipole due to the motion of the observer
with respect to the frame of statistical homogeneity and isotropy,
and the second, which can have arbitrary angular dependence,
is the corresponding term at the source. These are in addition
to the usual dipole due to the difference between the velocity
of the observer and the source.
As long as the difference between $u^\a$ and $n^\a$ is small,
the difference in the redshift between the two frames is small,
even though the expansion rates $\theta$ and $\theta^{(u)}$
can be very different, as the gradient of $v^\a$ can be
large even when $v$ is small.

\para{The local environment.}

When we argue for the cancellation of terms other than
$\av{\Gamma\theta}$ in the integral \re{Eint} due to symmetry,
this only applies to propagation over distances longer than the
homogeneity scale, and deviations due to the local environment
are not accounted for.
For example, in linearly perturbed FRW spacetimes, the shear term
$\sigma_{\a\b} e^\a e^\b$ contains the usual local dipole,
which we have neglected.
To be consistent in our approximation of concentrating on
propagation over long distances and neglecting the
effect of the local environments near the source and the observer,
we should approximate $1 - e^\a v_\a + v_\a e^\a|_0 \simeq 1$.
In the real universe, this approximation seems to hold well.
The velocity difference between the CMB frame and our
rest frame is of the order $10^{-3}$, and the rest frame
of local large-scale structures is also near the CMB frame \cite{dipole}.
The effect of the local environment is likely to be small
as long as structures are small compared to the
distance the light travels and the observer is not in a
special location \cite{Rasanen:2008a}.
This is true for the structures which are known to exist
and which are expected in usual models of structure formation,
but may not be valid for speculative large spherical structures,
often described with the Lema\^{\i}tre-Tolman-Bondi (LTB)
model \cite{LTB, February:2009}.

For the CMB anisotropies, the corrections
due to the local environment and the deviations around the mean
cannot be neglected. They are important
for the low multipoles, as in the Integrated Sachs-Wolfe
effect and the Rees-Sciama effect, and could be
related \cite{asymmodels} to observed violations of
statistical isotropy of the CMB \cite{asymobs}.
Formalism for the CMB in the case when the geometry is
not perturbatively near FRW has been developed in \cite{Maartens:1998}.

\para{The mean redshift and the scale factor.}

The redshift characterises a single geodesic (or more accurately,
two points and two frames along a single geodesic), so its spatial
average is not well defined. However, it is useful to introduce the 
``mean redshift'' $\bz$ by
\bea
  1 + \bz \equiv \exp\left( \int_{t}^{t_0}\rmd t \frac{1}{3} \av{ \Gamma \theta } \right) \ .
\eea

\noindent The physical interpretation of $1+\bz$ is that
if we take any two points on $\sN(t)$ and $\sN(t_0)$ which are
connected by a null geodesic (or several), the redshift
along the null geodesic(s) is $\bz$ plus small corrections
(assuming that the rest frames of the source and the observer
are close to the frame of statistical homogeneity and isotropy).
From the arguments above and the observational fact that the CMB
deviations from isotropy and from the blackbody shape of the
spectrum are small \cite{Fixsen:1996} we know that temperature
differences between different spatial locations are small,
and the mean value of the redshift gives the dominant contribution.

We define the scale factor $a$ as (setting $a(t_0)=1$)
\bea \label{a}
  a(t) &\equiv& (1 + \bz)^{-1} = \exp\left( - \int_{t}^{t_0}\rmd t \frac{1}{3} \av{ \Gamma \theta } \right) \ .
\eea

\noindent The quantity $\theta$ gives the change of rate of the
local volume element with respect to the proper time $s$, so
$\Gamma\theta$ gives the rate of change with respect to $t$.
Therefore $a(t)^3$ is proportional to the
volume of $\sN$: the mean redshift is determined by the
change of the overall volume of space.

The change of the redshift of a given source with time,
called redshift drift \cite{Sandage:1962}, has been suggested
as a test of the FRW metric and LTB
models \cite{Uzan:2008, Quartin:2009}. Essentially, the change of
redshift with time tests the relationship $1+z=a(t)^{-1}$ between
the redshift and the scale factor, with the scale factor
associated with an average expansion rate. In the present case,
unlike in LTB models, the relationship between
the mean redshift and the average expansion rate is the
same as in FRW models. (See \sec{sec:avexp} for discussion
of the average expansion rate.)
However, because redshift drift is a small effect,
the variations around the mean would have to be considered
carefully to make a prediction.

\subsection{Deflection} \label{sec:def}

\para{Picard's proof.}

In deriving \re{Eint} it was assumed that the
spatial direction of the null geodesic does not change
rapidly along the geodesic.
The direction $e^\a$ is the direction of the null geodesic
projected on $\sN$, so it enters into the arguments
about cancellation due to symmetry.
The direction $r^\a$ in turn describes the apparent position
of the source as seen by the observer, so it is an observable,
and the change in $r^\a$ is called the deflection.
As we do not have information about the 'original'
position of the source, i.e. the position in the hypothetical
situation that the spacetime would be flat along the photon path,
the deflection can only be measured statistically, unless the
apparent position changes on the timescale of the observation,
as in microlensing.

Let us look at the change of $e^\a$ and $r^\a$ along the
null geodesic. As with the redshift, it is simpler to
consider $e^\a$ and then relate it to $r^\a$.
It is convenient to choose a non-coordinate basis and introduce
tetrads adapted to the 1+3 decomposition
\cite{Ellis:1967, Ellis:1998c, vanElst:1996}.
We denote the tetrad basis by ${t^\a_{\ A}}$, with
$\eta^{AB} t_{\a A} t_{\b B}=g_{\a\b}, t^\a_{\ A} t_{\a B}=\eta_{AB}$
as usual, where $\eta_{AB}=\diag(-1,1,1,1)$.
We use capital Latin letters to denote components in the tetrad
basis, e.g. $e^A\equiv e^\a t_{\a}^{\ A}$.

The change of $e^A$ along the null geodesic is given by
\bea \label{eeq1}
  {E^{(n)}}^{-1} \patl{e^A} &=& ( n^B + e^B ) \nabla_B e^A \el
  &=& n^A \left( \frac{1}{3} \theta + \ndot_B e^B + \sigma_{BC} e^B e^C \right) - \ndot^{A} \el
  && + e^A ( \ndot_B e^B + \sigma_{BC} e^B e^C ) - \sigma^A_{\ \, B} e^B \ .
\eea

\noindent where we have on the second line inserted
$e^A= {E^{(n)}}^{-1} k^A - n^A$ inside the covariant derivative
and used \re{gradn} and \re{Eder}.
We specialise the choice of basis by taking $t_\a^{\ 0}=n_\a$,
so that $n^A=\delta^{A0}, n_A=-\delta_{A0}$.
(In a coordinate basis, this choice is not
possible in general \cite{Ehlers:1961, Ellis:1971}.)
Then $e^A$ is zero for $A=0$, while for the
spatial components (which we denote by small Latin letters from
the middle of the alphabet, $e^i\equiv h^i_{\ A} e^A$)
we obtain, using the definition of the covariant derivative,
\bea \label{grade}
  {E^{(n)}}^{-1} \patl{e^i} &=& ( n^B + e^B ) \nabla_B e^i \el
  &=& \pate{e^i} + ( n^B + e^B ) \Gamma^i_{\ CB} e^C \el
  &=& \pate{e^i} + a^i + \Omega^i_{\ j} e^j - a_j e^j e^i - \epsilon^{i}_{\ j k} N^j_{\ l} e^k e^l \ ,
\eea

\noindent where the connection components $\Gamma^A_{\ BC}$
have been expressed in terms of the decomposition \re{gradn}
of $\nabla_B n_A$ as well as an object $a_i$, a symmetric object $N_{ij}$
and an antisymmetric object $\Omega_{ij}$; see
\cite{vanElst:1996, Ellis:1998c} for details\footnote{In the
notation of \cite{vanElst:1996, Ellis:1998c},
$\Omega_{ij}\equiv\epsilon_{ijk}\Omega^k$.}.
Together with the 9 degrees of freedom
in $\{ \theta, \ndot^i, \sigma^i_{\ j} \}$, the 12 degrees
of freedom in $\{ a^i, N^i_{\ j}, \Omega^i_{\ j} \}$ completely
characterise the spacetime geometry.

Putting \re{eeq1} and \re{grade} together, we have a system
of ordinary differential equations for the components $e^i$
\bea \label{eeq}
  \pat_\eta e^i &=& - \ndot^i - a^i - ( \sigma^i_{\ j} + \Omega^i_{\ j} ) e^j + ( \ndot_j e^j + a_j e^j + \sigma_{jk} e^j e^k ) e^i + \epsilon^{i}_{\ j k} N^j_{\ l} e^k e^l \el
  &\equiv& f^i(\eta, e^j) \ .
\eea

\noindent If the geometry is statistically homogeneous and
isotropic and its distribution evolves slowly, the change in
$e^i$ remains small due to the lack of preferred directions in
$\{\ndot^i, a^i, \sigma^i_{\ j}, N^i_{\ j}, \Omega^i_{\ j} \}$.
This can be expressed more formally as follows.
According to Picard's theorem, the system of equations \re{eeq}
has a unique solution given by an iteration
(see e.g. \cite{Duff:1966}, page 19). Let us define
$e^i_{(N+1)}(\eta)\equiv e^i_{(0)} + \int_{\eta_0}^\eta \rmd\eta' f^i [\eta',e^i_{(N)}(\eta')]$,
with $e^i_{(0)}\equiv e^i(\eta_0)$. The solution
to \re{eeq} is given by the $N\rightarrow\infty$ limit.
At first step, we have
\bea \label{esol1}
  e^i_{(1)} &=& e^i_{(0)} - \int_{\eta_0}^\eta \rmd\eta' ( \ndot^i + a^i ) -  e_{(0)}^j \int_{\eta_0}^\eta \rmd\eta'( \sigma^i_{\ j} + \Omega^i_{\ j} ) + e_{(0)}^i e_{(0)}^j \int_{\eta_0}^\eta \rmd\eta' ( \ndot_j + a_j ) \el
  && + e^i_{(0)} e_{(0)}^j e_{(0)}^k \int_{\eta_0}^\eta \rmd\eta' \sigma_{jk}  + e_{(0)}^k e_{(0)}^l \int_{\eta_0}^\eta \rmd\eta' \epsilon^{i}_{\  j k} N^j_{\ l} \ .
\eea

\noindent As with \re{Eint}, we could write $\rmd\eta=\Gamma\rmd t$,
with the understanding that the spatial coordinates are functions
of $t$ along the null geodesic.
If $\ndot^i$ and $a^i$ have no preferred direction
and evolve slowly, their integral vanishes,
provided the distance over which the integral is taken
is longer than the homogeneity scale.
In practice, the cancellation is not perfect, because
there is evolution and statistical fluctuations,
so the integral is simply suppressed.
Similarly, all diagonal components of any tensor
should contribute almost equally to the integrals
and the contribution from non-diagonal components should
be suppressed.
Because $\sigma^i_{\ j}, \Omega^i_{\ j}$
and $\epsilon^{i}_{\ j k} N^j_{\ l}$ are traceless, their
contributions are suppressed.
One can repeat the argument at every iteration to conclude
that the solution $e^i$ is $e^i_{(0)}$ plus a small deviation.
(For the linear term in \re{eeq} this is obvious,
as it simply gives the $\eta$-ordered exponential of the
integral of $\sigma^i_{\ j} + \Omega^i_{\ j}$.)
The change in the observed position of the source
$r^A$ is then also small as long as $v$ is small, as
we see from \re{re}.
This qualitative understanding should be made more
rigorous, and the amplitude and the distribution of
the deflection should be evaluated.

\subsection{Null shear and angular diameter distance} \label{sec:da}

\para{The null shear.}

The distortion of the size and shape of the source
image are described by the null expansion rate $\thetat$
(or equivalently the angular diameter distance $D_A$)
and the null shear tensor $\sigmat_{\a\b}$.
To find these quantities, we need
to introduce a tensor $\htt_{\a\b}$ which projects onto
a two-space orthogonal to the null geodesic, $\htt_{\a\b} k^\b=0$.
Analogously to \re{gradn} and \re{gradu}, the covariant
derivative of $k^\a$ can be decomposed as follows:
\bea \label{kgrad}
  \nabla_\b k_\a &=& \thetat_{\a\b} \el
  &=& \frac{1}{2} \htt_{\a\b} \thetat + \sigmat_{\a\b} + k_{(\a} P_{\b)} \ ,
\eea

\noindent where the trace
$\thetat=\htt^{\a}_{\ \b} \nabla_\a k^\b=\nabla_\a k^\a$
is the expansion rate of the area of the null geodesic bundle,
$\sigmat_{\a\b}= \htt_{\a}^{\ \d} \htt_{\b}^{\ \c} \nabla_\c k_\d - \frac{1}{2} \htt_{\a\b} \thetat$
is the null shear and $P_\a$ is a vector which depends on
the choice of $\htt_{\a\b}$ and plays no role in what follows. 
We have $\sigmat_{\a\b} k^\b=0$, $P_\a k^\a=0$.
The null geodesic vorticity is zero,
because $k^\a$ is a gradient.
The null shear scalar is defined as
$\sigmat^2\equiv\ha\sigmat_{AB}\sigmat^{AB}$.
The area expansion rate $\thetat$ is related to the
angular diameter distance by (see e.g. \cite{Schneider:1992, Sasaki:1993})
\bea \label{DA}
  D_A \propto \exp \left( {\frac{1}{2} \int \rmd\l \thetat} \right) \ .
\eea

\noindent The angular diameter distance $D_A$ and the null
shear scalar $\sigmat^2$ are independent of $\htt_{\a\b}$.
However, $D_A$ depends on the observer velocity, because the
angular element is observer-dependent.
We consider $D_A$ in the $n^\a$ frame.
The observed angular diameter distance is then
\cite{Schneider:1992} (page 110)
$\frac{E^{(n)}}{E^{(u)}} D_A=\frac{1}{ \gamma ( 1 - v_\a e^\a ) } D_A
\simeq ( 1 + v_\a e^\a ) D_A$, where we have used \re{EuEn}.

It is again convenient to use tetrads instead of sticking to
a coordinate basis. We choose $\htt_{AB}$ to be orthogonal
to both $n^A$ and $e^A$,
\bea \label{htt}
  \htt_{AB} &=& g_{AB} + n_{A} n_{B} - e_{A} e_{B} \el
  &=& g_{AB} - {E^{(n)}}^{-2} k_{A} k_{B} + 2 {E^{(n)}}^{-1} k_{(A} n_{B)} \ ,
\eea

\noindent We proceed with $\sigmat_{AB}$ the same way
as we did with $e^A$ in \sec{sec:def}. Taking the derivative
along the null geodesic and projecting, we obtain
(see e.g. \cite{Schneider:1992, Sasaki:1993})
\bea \label{shear1}
  \htt_{(A}^{\ \ C} \htt_{B)}^{\ \ D} \patl{\sigmat_{CD}} = - \thetat \sigmat_{AB} - C_{AB} \ ,
\eea

\noindent where 
\bea
  C_{AB} &\equiv& k^M k^N \htt_{(A}^{\ \ C} \htt_{B)}^{\ \ D} C_{MCND} \el
  &=& 2 {E^{(n)}}^2 \htt_{(A}^{\ \ C} \htt_{B)}^{\ \ D} \left( E_{CD} + \ha \htt_{CD} e^M e^N E_{MN} - \tilde\epsilon_{CM} H^{M} _{\ \ D} \right) \,
\eea

\noindent where $C_{AB}$ has been expressed in terms
of the electric and magnetic components of the Weyl tensor,
$E_{AB} \equiv C_{ACBD} n^C n^D=C_{A0B0}$,
$H_{AB} \equiv \frac{1}{2} \epsilon_A^{\ \ CD} C_{CDBE} n^E=\frac{1}{2} \epsilon_A^{\ \ CD} C_{CDB0}$, and $\epst_{AB}\equiv\epsilon_{ABC} e^C$.

On the other hand, from the definition of the covariant derivative
we obtain (see \cite{Ellis:1998c, vanElst:1996} for the
decomposition of $\Gamma^A_{\ BC}$)
\bea \label{shear2}
  {E^{(n)}}^{-1} \htt_{(A}^{\ \ C} \htt_{B)}^{\ \ D} \patl{\sigmat_{CD}} &=& \pate \sigmat_{AB} - 2 \htt_{(A}^{\ \ C} \htt_{B)}^{\ \ D} ( n^E + e^E ) \Gamma^F_{\ C E} \sigmat_{FD} \el
  &=& \pate \sigmat_{AB} + \left( 2 \Omega_{ij} + 2 \epst_{k[i} N_{j]}^{\ k} + \epsilon_{ijk} N^{k}_{\ l} e^l \right) \htt_{(A}^{\ \ \, i} \sigmat_{B)}^{\ \ \, j} \ .
\eea

\noindent Note that in a tetrad basis contracting with $\htt_{AB}$
commutes with $\pate{}$, but not with $\patl{}$.
We denote indices on the space orthogonal to $n^A$
and $e^A$ with small Latin letters from the beginning of the
alphabet, $e^a\equiv \htt^a_{\ B} e^B$, and specialise the choice
of basis by taking $t_{\a}^{\ 3}=e_\a$, so that $e^A=\delta^{A3}, e_A=\delta_{A3}$.
Putting together \re{shear1} and \re{shear2}, we have
\bea \label{patsigma2}
  \pate \sigmat_{ab} &=&  - {E^{(n)}}^{-1} \thetat \sigmat_{ab} - {E^{(n)}}^{-1} C_{ab} + ( 2 \Omega_{c(a} +  2 \epst_{d[c} N^{d}_{\ (a]} + \epst_{c(a} N^{3}_{\ 3} ) \sigmat_{b)}^{\ \, c} \ .
\eea

Like equation \re{eeq} for the components of the deflection,
\re{patsigma2} is an ordinary differential equation for the two
independent components of $\sigmat_{ab}$. However, we cannot
straightforwardly apply Picard's theorem. First, \re{patsigma2}
contains the unknown $\thetat$.
Even if we include the equation \re{Raynull} given below
to obtain a closed first order system of equations, Picard's theorem does
not apply, because it assumes that the variables remain bounded,
whereas the initial condition for the area expansion rate is
$\thetat(\eta_0)=-\infty$ at the observer.

Nevertheless, the reasoning about cancellations due the lack
of preferred directions still holds, because the solution
depends on the source term $C_{ab}$ only via an integral.
This is transparent with the change of variable
$\sigmat_{ab}\equiv\tilde\Sigma_{ab}-\int\rmd\eta {E^{(n)}}^{-1} C_{ab}$.
We now argue as before that due to statistical homogeneity
and isotropy $C_{ab}$ contributes dominantly via its trace,
which is zero, so $\sigmat_{ab}\approx\tilde\Sigma_{ab}$.
This eliminates the source term in \re{patsigma2}.
Given the initial condition $\tilde\Sigma_{ab}(\eta_0)=0$,
we obtain $\sigmat_{ab}\approx\tilde\Sigma_{ab}\approx0$.
As with the deflection, this argumentation needs to
be made more rigorous, and the amplitude of the small
shear that is generated should be calculated.

\para{The angular diameter distance.}

Applying the derivative $\patl{}$ to $\thetat$, we
obtain the evolution equation for the area expansion rate
(see e.g. \cite{Schneider:1992, Sasaki:1993})
\bea \label{Raynull}
  \patl\thetat &=& - G_{AB} k^A k^B - 2 \sigmat^2 - \frac{1}{2} \thetat^2 \el
  &=& - 8\pi\GN T_{AB} k^A k^B - 2 \sigmat^2 - \frac{1}{2} \thetat^2  \el
  &=& - 8\pi\GN \big( \rho^{(n)} + p^{(n)} - 2 q^{(n)}_A e^A + \pi^{(n)}_{AB} e^A e^B \big) {E^{(n)}}^2 - 2 \sigmat^2 - \frac{1}{2} \thetat^2 \ ,
\eea

\noindent where we have used the Einstein equation
\re{Einstein} and applied the decomposition \re{emdecn}
of the energy-momentum tensor.
As discussed in \sec{sec:geom}, we could equally
regard \re{emdecn} as the decomposition of the
Einstein tensor (or, in the present context, the
Ricci tensor, as the trace does not contribute to \re{Raynull}).

Using \re{DA}, we obtain from \re{Raynull} the evolution
equation for the angular diameter distance:
\bea \label{DAeq}
  \frac{\rmd^2 D_A}{\rmd\l^2} &=& - \left[ 4 \pi\GN \big( \rho^{(n)} + p^{(n)} - 2 q^{(n)}_A e^A + \pi^{(n)}_{AB} e^A e^B \big) {E^{(n)}}^2 + \sigmat^2 \right] D_A \ .
\eea

The solution again depends on the source functions only via
an integral. This is transparent with the change of variable
$\thetat\equiv\tilde\Theta - 8\pi\GN \int\rmd\l ( \rho^{(n)} + p^{(n)} - 2 q^{(n)}_A e^A + \pi^{(n)}_{AB} e^A e^B ) {E^{(n)}}^2$.
We can write
\bea \label{muint}
  && \int\rmd\l ( \rho^{(n)} + p^{(n)} - 2 q^{(n)}_A e^A + \pi^{(n)}_{AB} e^A e^B ) {E^{(n)}}^2 \el
  && \approx E^{(n)}(\eta_0) \int\rmd t a^{-1} \av{ \Gamma ( \rho^{(n)} + p^{(n)} ) } \el
  && = E^{(n)}(\eta_0) \int\rmd t a^{-1} \av{ \Gamma ( \rho^{(u)} + p^{(u)} )+ \frac{4}{3} \Gamma F } \ ,
\eea

\noindent where 
$F\equiv v^2 (\rho^{(u)} + p^{(u)}) + 2 \gamma^{-1} q^{(u)}_A v^A + \pi^{(u)}_{AB} v^A v^B\simeq 2 q^{(u)}_A v^A $.
On the second line we have taken into account
$\rmd\l={E^{(n)}}^{-1}\rmd\eta$ and the approximate scaling
$E^{(n)}\approx E^{(n)}(\eta_0) a^{-1}$.
We have also again applied the reasoning that statistical
homogeneity and isotropy together with slow evolution implies
that the contributions of $q^{(n)}_A e^A$ and $\pi^{(n)}_{AB} e^A e^B$
are suppressed, and that the dominant
contribution of $\rho^{(n)} + p^{(n)}$ comes from the average,
Finally, we have written $\rho^{(n)} + p^{(n)}$ in terms of $u^\a$
frame quantities using \re{h}, \re{n}, \re{emdecn} and \re{emdecu}.
Dropping the null shear, the solution $\thetat$ to \re{Raynull}
depends only on the quantity \re{muint} and
$\l=\int\rmd\eta {E^{(n)}}^{-1} \approx E^{(n)}(\eta_0)^{-1}\int\rmd t a\av{\Gamma}$,
where we have assumed that $\Gamma$ has a statistically
homogeneous and isotropic distribution and varies slowly, so
that the integral is dominated by the average value.
Because both \re{muint} and $\l$ depend approximately only
on $t$ (and $E^{(n)}(\eta_0)$), so does $\thetat$.
Writing \re{Raynull} as an integral equation, dropping
subdominant parts which depend on position, taking the
time derivative and expressing the equation in terms
of the angular diameter distance, we obtain
\bea \label{DAbareq}
  H \pat_{\bz} \left[ \av{\Gamma}^{-1} (1+\bz)^2 H \pat_{\bz} \bar{D}_A \right] &=& - 4\pi\GN \av{ \Gamma (\rho^{(u)} + p^{(u)}) + \frac{4}{3} \Gamma F } \bar{D}_A \el
  &\simeq& - 4\pi\GN \av{ \Gamma (\rho^{(u)} + p^{(u)}) + \frac{8}{3} \Gamma q^{(u)}_A v^A } \bar{D}_A \ ,
\eea

\noindent where the notation $\bar{D}_A(t)$ refers to the dominant
part of the angular diameter distance with the corrections to the
mean dropped, and we have used the relation
$\pat_t\bar{D}_A=-(1+\bz) H \pat_{\bz}\bar{D}_A$, with $H\equiv\adot/a$.
The quantity $\bar{D}_A$ has a similar physical interpretation
as $\bz$: if we take any two points on $\sN(t)$ and $\sN(t_0)$
connected by a null geodesic (or several of them), the angular
diameter distance along the geodesic(s) is $\bar{D}_A$
plus small corrections.
As noted in \cite{Rasanen:2008b}, while $\bar{D}_A(\bz)$
is well-defined, there does not exist a function $D_A(z)$ even along
a single geodesic, because the redshift is in general
not monotonic along the null geodesic\footnote{In
\cite{Rasanen:2008b} it was incorrectly claimed that when
the factor in the parenthesis on the right-hand side of
\re{DAeq} is positive, $D_A$ would be monotonic, because
the initial condition (at the observer) for $\patl{D_A}$
is negative. However, because $\l$ decreases
along the null geodesic away from the observer, this
only implies that $\patl{D_A}$ has at most one zero,
so $D_A$ is separately monotonic on at most two sections
of the null geodesic, as is well known from FRW spacetimes.
In the present case, the sign of the factor in the parenthesis
is not determined, and the number of zeros of $\patl{D_A}$ is not limited.}.

Apart from $\Gamma$ and $F$, the equation \re{DAbareq}
for the mean distance in terms of $\bz$ is the same as in
FRW spacetimes.
However, the relation between $\av{\rho^{(u)} + p^{(u)}}$ and
$\bz$ is different than in the FRW case, as we discuss
in the next section (see also \cite{Rasanen:2008b}).
As with the redshift and the
deflection, it is important to make the arguments
about the null shear and the angular diameter distance
more rigorous by evaluating the variation around the
mean. Observationally, the angular diameter distance
is known not to vary much with direction \cite{peakvar}.

\section{The average expansion rate and the scale factor} \label{sec:av}

\subsection{The average expansion rate} \label{sec:avexp}

\para{Defining the average.}

We have started with light propagation, which involves
the observed quantities directly, and have been led to
consider averages. The average of a scalar $f$ on $\sN$ is
\bea \label{av}
  \av{f}(t) \equiv \frac{ \int \epsilon f }{ \int \epsilon } \ ,
\eea

\noindent where $t$ is constant on $\sN$.
Recall that in general, $t$ is not a proper time.
In particular, it is neither the proper time
associated with $n^\a$ nor the proper time measured
by the observers. The commutation rule between averaging
and taking a derivative with respect to $t$ is
\bea \label{comm}
  \pat_t\av{f} = \av{\pat_t f} + \av{\Gamma\theta f} - \av{f} \av{\Gamma\theta}
\eea

The scale factor $a$ was defined in \re{a} to give the mean
redshift. From the definition it follows that $a^3$ is
proportional to the volume of $\sN$. The average expansion
rate of interest is
\bea \label{H}
  3 \H &=& \av{\Gamma\theta} \el
  &=& \av{ \Gamma \gamma^{-1} \theta^{(u)} - \path_\a ( \Gamma v^\a ) - \Gamma \gamma^2 (\vdot_\a v^\a + v^\a v^\b \path_\a v_\b) } \el
  &\simeq& \av{ \Gamma \theta^{(u)} - \path_\a ( \Gamma v^\a ) - \Gamma \vdot_\a v^\a } \ ,
\eea

\noindent where we have used \re{reltheta} and the relation
$\ndot_\a=\Gamma^{-1}\path_\a\Gamma$.
In the irrotational ideal fluid case, the scale factor
was originally defined as the volume of the hypersurface
orthogonal to $u^\a$, or equivalently by using the
average of $\theta^{(u)}$, with the lapse function
included \cite{Buchert:1999, Buchert:2001}.
We start from light propagation, and while $u^\a$
is the relevant velocity field for the redshift, the
symmetry of the spacetime selects $\Gamma \nabla_\a n^\a=\Gamma\theta$
as the relevant local expansion rate.
Locally $\theta$ can be very different from $\theta^{(u)}$, even for
small $v$, because the derivatives of $v^\a$ can be large.
However, using Gauss' theorem \cite{Wald:1984} (page 433)
the total derivative in \re{H} can be converted into a surface
integral which describes the flux of $\Gamma v^\a$ through the boundary.
If the distribution is statistically homogeneous and isotropic,
there should be an equal flux through the surface in both
directions, so the integral vanishes (up to statistical
fluctuations).
Therefore, the difference between the average quantities
$\av{\Gamma\theta}$ and $\av{\Gamma\theta^{(u)}}$ is suppressed
by $v$.

\subsection{Evolution equations for the scale factor} \label{sec:evo}

\para{The average equations.}

We now write down the evolution equations for
the scale factor $a$, or equivalently for the
average expansion rate.
These generalise the Buchert equations derived
for irrotational dust \cite{Buchert:1999}.
In \cite{Buchert:2001} the average equations were
written down in the irrotational ideal fluid case
using the ADM formalism, assuming that the
averaging hypersurface is orthogonal to the fluid flow\footnote{These
average equations, apart from the conservation law
of the energy-momentum tensor, were written down in the context of
general irrotational matter content in \cite{Behrend}.
Note that the perturbative calculation in \cite{Behrend} is incorrect,
because the averages of both first order terms and intrinsic second order
terms taken in the perturbed spacetime are neglected in comparison
with the averages of squares of first order terms. In fact,
all these terms are of the same order, and the distinction between
them is gauge-dependent \cite{Kolb:2004}.}.
In \cite{Larena:2009} the average equations were derived
(also in the ADM formalism) for an ideal fluid
including rotation, taking the expansion rate to be
$h^{\a\b}\nabla_\b u_\a=\theta^{(u)}+n_\a \dot u^\a$,
and keeping the hypersurface of averaging arbitrary.
(The formalism was applied to second order perturbation
theory in \cite{Clarkson:2009a}, with the hypersurface
fixed by the condition $H_{\a\b}=0$.
Averaging in second order perturbation theory was also
considered in \cite{Brown:2009}, with different hypersurfaces
fixed by coordinate conditions.)
In \cite{Gasperini:2009}, the average equations were
derived for an ideal fluid in the covariant formalism,
with an arbitrary averaging hypersurface.
We consider general matter content, an arbitrary
hypersurface of averaging and use the covariant formalism.

Combining the Einstein equation \re{Einstein}
with the Bianchi and Ricci identities for $n^\a$,
the evolution equations can be conveniently written in
terms of the decompositions \re{gradn} and \re{emdecn}
and the electric and magnetic components of the Weyl tensor
\cite{Ellis:1998c, Clarkson:2000, Tsagas:2007}.
We are only interested in the three scalar equations
\bea
   \label{Rayloc} \thetadot + \frac{1}{3} \theta^2 &=& - 4\pi\GN ( \rho^{(n)} + 3 p^{(n)} ) - 2 \sigma^2 + \ndot_\a \ndot^\a + \path_\a \ndot^\a \\
  \label{Hamloc} \frac{1}{3} \theta^2 &=& 8 \pi \GN \rho^{(n)} - \frac{1}{2} \sR + \sigma^2 \\
  \label{consloc} \rhodot^{(n)} + \theta ( \rho^{(n)} + p^{(n)} ) &=& - \path_\a q^{(n)\a} - 2 \ndot_\a q^{(n)\a} - \sigma_{\a\b} \pi^{(n)\a\b} \ .
\eea

\noindent where $\sR$ is the spatial curvature of $\sN$.
If $\sN$ is a hypersurface of constant proper time,
then $\ndot^\a=0$, and the equations differ from
the irrotational dust case only by the pressure term in the
Raychaudhuri equation \re{Rayloc} and the non-dust terms in
the conservation law \re{consloc}. In terms of the Hamiltonian
constraint \re{Hamloc}, the only difference is the different
evolution of the energy density. We keep $\ndot^\a$ arbitrary.
Changing to derivatives with respect to $t$, averaging,
applying the commutation rule \re{comm} and using
the relations $\av{\Gamma\theta}=3\adot/a$ and
$\ndot_\a=\Gamma^{-1}\path_\a\Gamma$, we obtain
\bea
  \label{Rayavn} && 3 \frac{\addot}{a} = - 4\pi\GN \av{\rho^{(n)} + 3 p^{(n)}} + \av{\ndot_\a \ndot^\a} + \av{\path_\a \ndot^{\a}} + \sQ \el
  && \quad + \av{ \frac{1}{3} (\Gamma^2-1) \theta^2 + (1-\Gamma^{-2})\Gamma \pat_t\theta + \theta \pat_t\Gamma } \\
  \label{Hamavn} && 3 \HH = 8 \pi \GN \av{\rho^{(n)}} - \ha \av{\sR} - \ha \sQ + \frac{1}{3} \av{(\Gamma^2-1)\theta^2} \\
  \label{consavn} && \pat_t \av{\rho^{(n)}} + 3 \H \av{\rho^{(n)} + p^{(n)}} = - \av{\Gamma\theta p^{(n)}} + \av{\Gamma\theta} \av{p^{(n)}} - \av{\Gamma \ndot_\a q^{(n)\a} + \Gamma \sigma_{\a\b} \pi^{(n)\a\b}} \el
  && \quad - \av{\path_\a (\Gamma q^{(n)\a})} \ ,
\eea

\noindent where
\bea
  \sQ &\equiv& \frac{2}{3}\left( \av{{\Gamma\theta}^2} - \av{\Gamma\theta}^2 \right) - 2 \av{\sigma^2} \ .
\eea

\noindent In \cite{Buchert:2001, Behrend, Larena:2009},
the equivalent of the factors of $\Gamma$ were put inside the
averages of the terms which appear on the right-hand side of
\re{Rayloc} and \re{Hamloc}, rather than the left-hand side.
Inserting \re{Rayloc} into the last term of \re{Rayavn} and
\re{Hamloc} into the last term of \re{Hamavn} would recover
that form of the equations.
However, the present convention keeps the $\path_\a\ndot^\a$
term and the last term of \re{consavn} as total derivatives,
which we can neglect.

Because the backreaction variable $\sQ$ is a statistical
quantity which characterises the distribution of the
spatial geometry, it is appropriate to give it in
terms of $\theta$ and $\sigma$, which are related to $n^\a$.
However, it seems more appropriate to express the energy-momentum tensor
in terms of the decomposition with respect to $u^\a$ from the point of
view of estimating the magnitude of the different terms.
Using \re{h}, \re{n}, \re{emdecn} and \re{emdecu}, we obtain
\bea
  \label{Rayavu} && 3 \frac{\addot}{a} = - 4 \pi\GN \av{\rho^{(u)} + 3 p^{(u)}} + \av{\ndot_\a \ndot^\a} + \av{\path_\a \ndot^{\a}} + \sQ \el
  && \quad + \av{ \frac{1}{3} (\Gamma^2-1) \theta^2 + (1-\Gamma^{-2})\Gamma \pat_t\theta + \theta \pat_t\Gamma } - 8 \pi\GN \av{F} \\
  \label{Hamavu} && 3 \HH = 8 \pi \GN \av{\rho^{(u)}} - \ha \av{\sR} - \ha \sQ + \frac{1}{3} \av{(\Gamma^2-1)\theta^2} + 8 \pi\GN \av{F} \\
  \label{consavu} && \pat_t \av{\rho^{(u)}} + 3 \H \av{\rho^{(u)} + p^{(u)}} = - \av{\Gamma\theta ( p^{(u)} + \frac{1}{3} F ) } + \av{\Gamma\theta} \av{p^{(u)} + \frac{1}{3} F} \el
  && \quad - \av{\gamma \Gamma \ndot_\a q^{(u)\a} + \gamma^2 \Gamma
    (\rho^{(u)}+p^{(u)}) \ndot_\a v^\a + \gamma \Gamma \ndot_\a
    q^{(u)}_\b v^\a v^\b + \Gamma \ndot_\a \pi^{(u)\a\b} v_\b } \el 
  && \quad - \av{ \Gamma \sigma_{\a\b} \pi^{(u)\a\b} + \gamma^2 \Gamma (\rho^{(u)}+p^{(u)}) \sigma_{\a\b} v^\a v^\b + 2 \gamma \Gamma \sigma_{\a\b} q^{(u)\a} v^\b } \el
  && \quad - \pat_t \av{F} - 4 \H \av{F} - \av{\path_\a (\Gamma q^{(n)\a})} \ ,
\eea

\noindent with $F=v^2 (\rho^{(u)} + p^{(u)}) + 2 \gamma^{-1}
q^{(u)}_\a v^\a + \pi^{(u)}_{\a\b} v^\a v^\b$ as before.
We have not written $\path_\a (\Gamma q^{(n)\a})$ in terms of
$u^\a$ frame quantities, because it is suppressed due to
statistical homogeneity and isotropy, like $\path_\a \ndot^\a$.
Vorticity does not appear in the above equations, because
$n^\a$ is hypersurface-orthogonal by construction.
Were we to decompose $\nabla_\b n_\a$ with respect
to the $u^\a$ frame, the vorticity of $u^\a$ would
(to leading order in $v$) emerge from
$\path_\a \ndot^\a$ and $\sR$ \re{Rayloc} and \re{Hamloc}.
(For the definition of $\sR$ for velocity fields which are not
hypersurface-orthogonal, see \cite{Ellis:1990}.)
In particular, the leading order contribution of the $u^\a$
frame vorticity to the average Raychaudhuri equation \re{Rayavn}
vanishes due to statistical homogeneity and isotropy, because it
is contained in the boundary term $\av{\path_\a \ndot^\a}$.
(In Newtonian gravity, the vorticity combines with $\sQ$ to give
a boundary term, so backreaction vanishes for periodic
boundary conditions and for statistical homogeneity and
isotropy \cite{Buchert:1995}.)

Dropping the boundary terms $\path_\a (\Gamma q^{(n)\a})$ and
$\path_\a \ndot^\a$ as well as all terms higher than first
order in $v$, we have
\bea
  \label{Rayapp} && 3 \frac{\addot}{a} \simeq - 4 \pi\GN \av{\rho^{(u)} + 3 p^{(u)}} + \av{\ndot_\a \ndot^\a} + \sQ \el
  && \quad + \av{ \frac{1}{3} (\Gamma^2-1) \theta^2 + (1-\Gamma^{-2})\Gamma \pat_t\theta + \theta \pat_t\Gamma } - 16 \pi\GN \av{q^{(u)}_\a v^\a} \\
  \label{Hamapp} && 3 \HH \simeq 8 \pi \GN \av{\rho^{(u)}} - \ha \av{\sR} - \ha \sQ + \frac{1}{3} \av{(\Gamma^2-1)\theta^2} + 16 \pi\GN \av{q^{(u)}_\a v^\a} \\
  \label{consapp} && \pat_t \av{\rho^{(u)}} + 3 \H \av{\rho^{(u)} + p^{(u)}} \simeq - \av{\Gamma\theta \big( p^{(u)}+\frac{2}{3}q^{(u)}_\a v^\a \big)} + \av{\Gamma\theta} \av{ p^{(u)} + \frac{2}{3} q^{(u)}_\a v^\a } \el
  && \quad - \av{ \Gamma \ndot_\a q^{(u)\a} + \Gamma (\rho^{(u)}+p^{(u)}) \ndot_\a v^\a + \Gamma \ndot_\a v_\b \pi^{(u)\a\b} } - \av{ \Gamma\sigma_{\a\b} \pi^{(u)\a\b} + 2 \Gamma \sigma_{\a\b} q^{(u)\a} v^{\b} } \el
  && \quad - 2 \pat_t \av{q^{(u)}_\a v^\a} - 8 \H \av{q^{(u)}_\a v^\a} \ .
\eea

\para{The dust approximation.}

One reason for deriving the general equations
\re{Rayavu}--\re{consavu} is to take into account
deviations from the approximation of treating
the matter as dust in the late universe.
The importance of the different terms depends
on the matter model, and cannot be determined from
general arguments. However, it is possible
to say what would would be necessary for the
non-dust terms to have a significant effect.
For the $\ndot^\a$ terms to be important in \re{Rayapp}--\re{consapp},
$\ndot_\a \ndot^\a$ would have to be of the order of the
square of the expansion rate in a large fraction of space (contrary
to what was argued in \cite{Tsagas:2009}; see also \cite{Coley}),
or the contraction of the energy flux and $\ndot^\a$ would
have to be of the order of the product of the average energy
density and the average expansion rate.
In order for the pressure or the anisotropic stress
to be important, they would have to be on average of the same order
of magnitude as the average energy density.
For the time dilation to be important, the spatially
varying part of $\Gamma$ would have to be of order one
in a fraction of space which is of order one.
If the matter content is a gas of Standard Model particles
plus cold or warm dark matter, and structures evolve from
small adiabatic perturbations with a nearly scale-invariant
spectrum, it seems unlikely that any of these conditions would
be satisfied in the late universe when radiation pressure can be neglected.

Let us assume that the matter is approximately dust in
the $u^\a$ frame, i.e. that $p^{(u)}, q^{(u)\a}, \pi^{(u)}_{\a\b}$
and $A^\a$ are small\footnote{From the equations
of motion it follows that $A^\a$ is zero if
$p^{(u)}, q^{(u)\a}$ and $\pi^{(u)}_{\a\b}$ are zero, and $A^\a$ is small if
$p^{(u)}, q^{(u)\a}, \pi^{(u)}_{\a\b}$ as well as $\path_\a p^{(u)}$ and
$\path^\b \pi^{(u)}_{\a\b}$ are small.},
and the deviation of $\Gamma$ from unity (and the
time derivative of the deviation) is small,
$\Gamma\equiv1-\delta\Gamma$, with $|\delta\Gamma|\ll1$.
When we drop all squares of small terms (whether
they are non-dust terms or $v^\a$), the
equations \re{Rayapp}--\re{consapp} simplify to
\bea
  \label{Rayav} 3 \frac{\addot}{a} &\simeq& - 4 \pi\GN \av{\rho^{(u)} + 3 p^{(u)}} + \sQ \el
  && + \av{ \vdot_\a \vdot^\a + \frac{2}{3} \delta\Gamma \theta^2 + 2 \delta\Gamma\pat_t\theta + \theta \pat_t\delta\Gamma } \\
  \label{Hamav} 3 \HH &\simeq& 8 \pi \GN \av{\rho^{(u)}} - \ha \av{\sR} - \ha \sQ + \frac{2}{3} \av{ \delta\Gamma \theta^2} \\
  \label{consav} \pat_t \av{\rho^{(u)}} + 3 \H \av{\rho^{(u)}} &\simeq& - \av{\theta p^{(u)}} + \av{ \vdot_\a q^{(u)\a} + \rho^{(u)} \vdot_\a v^\a } - \av{\sigma^{(n)}_{\a\b} \pi^{(u)\a\b} } \ ,
\eea

\noindent where we have used the fact that
$\ndot^\a\simeq A^\a-\vdot^\a$ plus corrections of order $v$.
Equations \re{Rayav}--\re{consav} give the leading
corrections to the treatment of matter as irrotational dust,
compared to the original Buchert equations \cite{Buchert:1999}.
In particular, they cover the case when the matter can be
locally treated as dust, but has rotation, and $n^\a$ is orthogonal
to the hypersurface of constant proper time of observers
comoving with the dust. Then $\vdot^\a$ is of order $v$,
and we can choose $\Gamma=1$, so the difference between
$n^\a$ and $u^\a$ arises only from vorticity.
We see that vorticity alone has (to leading order in $v$)
no effect on the averages, because the dominant contribution
comes from the total derivative $\path_\a \ndot^\a$. 
It does change the relation of the scale factor $a$
to the geometry, because $a$ will be defined with a
different vector field, but not the relation of $a$ to the redshift.

For irrotational dust, the Raychaudhuri equation \re{Rayloc}
can be integrated as an inequality before averaging to obtain
the bound $Ht\leq1$ \cite{Wald:1984} (page 220),
\cite{Nakamura:1995, Rasanen:2005}.
When rotation or non-dust terms are important,
there is no such local inequality.
Indeed, having $\thetadot+\frac{1}{3}\theta^2>0$ locally
is required in order for collapsing regions to stabilise.
From a physical point of view, we would still expect
to recover $Ht\leq1$ unless there is sustained acceleration
in a significant fraction of space, but the conditions for this
derived from \re{Rayloc} involve combinations of spatial
averages and integrals over time, and are not entirely transparent.

Note that in order for the approximation of treating the matter
as dust on average to hold, it is only necessary that the contribution of
the non-dust terms to the averages is smaller than that of the average
energy density. It is not required that the energy-momentum
tensor of matter could be locally approximated as dust everywhere.
In fact, deviations from the irrotational dust behaviour
are necessarily important on small scales.
As the local Raychaudhuri equation \re{Rayloc}
shows, in order to stabilise structures, a large
$\ndot^\a$ or its gradient is needed.
This can correspond to $u^\a$ frame vorticity
as with rotating baryonic structures,
or the acceleration can be generated by
anisotropic stress (or a pressure gradient or energy flux)
as in the case of dark matter \cite{Pueblas:2008}.
For dark matter, the dust approximation is
locally invalid in structure formation due to
multistreaming \cite{Buchert:2005}.
Nevertheless, as long as the volume occupied
by regions where such terms are important is small, their
contribution to the average is not important.
In Newtonian calculations, this is certainly the case \cite{Pueblas:2008}.

Approximating the matter content as dust on average does not
imply  viewing the matter as infinitesimal grains.
For example, the issue of what the ``particles''
of the dust fluid are is sometimes raised, and whether one
can consistently ``renormalise'' the scale of the
description as larger stable structures form
\cite{Rasanen:2006b, Rasanen:2008a}.
However, the dust approximation is properly understood
as the statement that when considered on large scales,
the energy density dominates over the pressure, the energy
flux, the anisotropic stress, and their gradients.

It has been argued that because of gradients of
spatial curvature, clocks in different regions
run at different rates, and that this effect is important
for cosmology but neglected in the dust approximation \cite{Wiltshire}.
Any such effects are accounted for in the present analysis,
to the extent they are part of general relativity, and
not outside of the geometrical optics approximation.
If the $n^\a$ frame is non-geodesic, different points on the
hypersurface of statistical homogeneity and isotropy indeed
have different values of the $n^\a$ frame proper time
(though this cannot be understood as being due to spatial curvature
gradients), which in turn is close to the proper time
measured by the observers if $v$ is small.
For this time dilation to be significant, the acceleration
$\ndot^\a$ would have to be of the order of the average
expansion rate in a significant fraction of space.
This in turn requires that either the motion of the observers
is very non-geodesic (large $A^\a$) or the acceleration
between the two frames is significant (large $\vdot^\a$);
the latter possibility however has to contend with the fact
that the velocity between the frames cannot become large,
as this would violate the small anisotropy observed in the CMB.
In order to generate such large accelerations, the non-dust terms
in the energy-momentum tensor would have to be significant in a large
fraction of space. This would likely have important cosmological
effects apart from the time dilation.
Note that it follows from statistical homogeneity and isotropy
that the spatial difference in the CMB temperature
between different regions is small, in contrast to
the arguments made in \cite{Wiltshire}.
This issue can be observationally probed with the
blackbody shape of the CMB spectrum \cite{Fixsen:1996, blackbody}.

\section{Discussion}

\subsection{Modelling issues}

\para{Discreteness.}

While we have kept the energy-momentum tensor generic,
the arguments about light propagation in \sec{sec:light}
contain the implicit assumption that matter is so
finely distributed in space that it can be treated
as a continuous distribution which light rays sample.
The redshift, the angular diameter distance and other quantities
related to light propagation depend on the spacetime geometry
only via an integral along the null geodesic. If the matter
consists of discrete clumps whose size and number density is
so small that a typical light ray will never encounter matter,
the energy-momentum tensor along the light ray is zero, regardless
of the average energy density (or pressure or other components).
For example, the integrand in \re{muint} vanishes identically,
and our arguments about cancellation between regions of low
and high density do not apply.

The approximation of discrete matter as a continuous fluid
has been studied from first principles for the dynamics of
matter in Newtonian gravity \cite{discrete}.
However, the effect on light propagation has been looked
at mostly from the perspective of adding perturbations to
a FRW metric. With a statistically homogeneous and isotropic
distribution and small structures, it is then not surprising
that the deviation of the expansion rate from the FRW case
is small \cite{Rasanen:2008a}.
One exception is \cite{Clifton}, where the effect of discreteness
light propagation was considered in a lattice model without any FRW
approximation.

If the light travels in vacuum, and we assume statistical
homogeneity and isotropy, the mean angular diameter distance
is given by \re{DAbareq} with zero on the right-hand side.
The equation can be integrated to yield
\bea \label{DAempty}
  \bar{D}_A = \int_0^{\bz} \rmd z \frac{\bar{\Gamma}(z)}{(1+z)^2 \bar{H}(z)} \ ,
\eea

\noindent where $\bar{H}$ and $\bar{\Gamma}$ are the mean expansion rate
and the mean time dilation along the null geodesic,
which in general do not coincide with the spatial averages.
Because the null geodesic samples only vacuum, the expansion rate along
the null geodesic is larger than the average
expansion rate (assuming that the matter satisfies the
strong energy condition). The equations \re{Rayloc} and \re{Hamloc}
show that if the acceleration $\ndot^\a$ and the shear
$\sigma_{\a\b}$ could be neglected, the expansion rate sampled
by the light ray would be the same as in an empty FRW universe.
If the time dilation could be neglected, the angular diameter
distance would then correspond to the 'coasting universe'
(or Minkowski space, for non-expanding regions).
However, it is probably not reasonable to neglect
the shear $\sigma_{\a\b}$. For example, the existence of both expanding
and non-expanding regions means that there is a gradient
in the expansion rate, which implies non-zero shear.
Evaluating the expansion rate along a null geodesic,
and the angular diameter distance, is thus reduced to the
question of realistically modelling the shear scalar (and the
acceleration $\ndot^\a$) along the geodesic.

Discussing light propagation in terms of null geodesics assumes
the validity of the geometrical optics approximation.
Geometrical optics is in turn based on modelling
light as local plane waves, which requires the wavelength
of the light to be much smaller than both the curvature scale
and the scale over which the amplitude, wavelength and
polarisation of the light change significantly. If the fraction of the
volume occupied by matter is so small that light rays
never come close to the matter particles, this is satisfied.
However, if the light passes through small discrete
regions where the energy-momentum tensor is non-zero,
the situation is very different from the geometrical optics limit.
(We are here concerned only with gravitational
interactions, and are not taking into account gauge
interactions between photons and matter.)
In evaluating the validity of the continuous fluid
approximation, the extension of the wave-packet of the
matter particles should therefore be taken into account.
The treatment of the photon waves should also be more
detailed, instead of simply treating their transverse
width as zero (as in the null geodesic picture) or
infinite (as implicit in the plane wave approximation).
The effect of discreteness on light propagation is not
obvious, and should be studied in a realistic model.

\para{The relevance of averages.}

The equations \re{Rayavn}--\re{consavn} or
\re{Rayavu}--\re{consavu} generalise
the average equations for the irrotational
dust case derived by Buchert \cite{Buchert:1999}
to arbitrary matter content and rotation,
and an arbitrary averaging hypersurface.
The scale factor has been defined to give the mean
redshift in the case that the difference between the
observer frame and the frame of statistical homogeneity
and isotropy is small, and we have assumed statistical
homogeneity and isotropy and slow evolution, which are
required to have a meaningful notion of mean redshift.
The equations \re{Rayavn}--\re{consavn} or
\re{Rayavu}--\re{consavu} are of course valid
without any symmetry assumptions. However,
the quantity $a^3$ is of limited use in interpreting observations
unless the space is statistically homogeneous and isotropic,
so that the change in the total volume of the spatial hypersurface
is the dominant effect for light propagation.

The system of equations \re{Rayavn}--\re{consavn}
or \re{Rayavu}--\re{consavu} is derived from the
scalar part of the full Einstein equation, which is
not closed, because it is coupled to the vector and tensor parts.
As the sum of two tensors at different spacetime
points is not a tensor, the average of a tensor (or vector)
in curved spacetime is not well-defined, so it is sometimes
said that the rest of the evolution equations cannot be averaged.
However, we can write the evolution equations
in terms of components, and average these.
The problem is not lack of covariance\footnote{One can even average
the vector and tensor part of the equations covariantly
by first projecting with a vector field such as $v^\a$
or $\path_\a\theta$. The Einstein equation can be
expressed in full generality in scalar form
by using a projection \cite{Padmanabhan:2004}.},
but the fact that products of variables become
independent correlation terms, so the number of unknowns
increases when taking the average, and the set of average
equations does not close.

Methods for covariantly averaging tensors on curved spacetime
have been suggested, including the macroscopic gravity
formalism \cite{Zalaletdinov, Paranjape}, the Ricci flow \cite{Carfora},
a statistical averaging formalism \cite{Debbasch},
a procedure which relies on a specific choice of tetrads \cite{Behrend:2008},
and the proposal of \cite{Korzynski:2009} which is more
a way to rewrite the tensors than average them.
However, the relevant issue is not the mathematical definition
of averages in some covariant formalism which is an extension
of general relativity or in a statistical ensemble of spacetimes.
Rather, we want to determine the impact of structures in
the spacetime which actually describes the universe we
observe, with the dynamics determined by the Einstein equation
and the local equations for light propagation.
Averages are useful insofar as they provide an approximate
description of observed quantities in this complex system.
The relevant averaging procedure, and the hypersurface
of averaging, emerges from considering observations,
and cannot be determined on abstract mathematical grounds. 

It bears emphasising that taking the average on a different
hypersurface would correspond to considering a different velocity
field, and this is a physical choice. This issue is
separate from the question of gauge-invariance, i.e. dependence of
unphysical quantities on the chosen coordinate system.
The Buchert equations were originally derived using the ADM
formalism \cite{Buchert:1999}, where the distinction between
choice of velocity field and choice of coordinates is
not entirely transparent. However, the problem can be considered
completely covariantly, without introducing coordinates
\cite{Tsagas:2007, Rasanen:2008a, Rasanen:2008b}.
The averages depend on the choice of the averaging hypersurface
\cite{Geshnizjani, Rasanen:2004}, but not on the coordinate
system \cite{Kolb:2004}\footnote{In \cite{Brown:2009},
choice of the averaging hypersurface and the coordinate system
was conflated. This mixes up defining a quantity of interest
using physical criteria and using different coordinates to
describe the same physics.}.
The relevant velocity field is singled out as that of the
observer by the redshift, and the relevant averaging hypersurface
is given by the symmetry properties of the spacetime.

\para{Deviation from the FRW universe.}

If backreaction is important for the average
expansion rate, i.e. if $\sQ$ contributes significantly
to \re{Rayavn} and \re{Rayavu}, there is no ``FRW background''
that would emerge on large scales \cite{Rasanen:2006b, Rasanen:2008a}.
(Note that $\sQ$ being small does not guarantee that the spacetime
can be described by the FRW metric.)
The FRW metric describes a universe that is exactly
homogeneous and isotropic, not a universe where there
are large non-linearities with a statistically homogeneous
and isotropic distribution.

While the deviation of the average expansion rate from the
FRW equations could be attributed to an effective matter
component in a FRW universe, this is not the case for
other observables. For the shear scalar, this is obvious,
because it is zero in FRW models, but generally positive.
As a less trivial example, the spatial curvature 
in the FRW case is fixed to be proportional to $a^{-2}$, while the
average spatial curvature in an inhomogeneous and/or
anisotropic space can evolve non-trivially \cite{Rasanen:2007}.
In fact, if the
matter can be treated as dust, the effect of backreaction
is encoded in the non-trivial evolution of the spatial curvature
\cite{Buchert:1999, Rasanen:2005, Rasanen:2006b}. (The
difference between backreaction in general relativity
and Newtonian gravity can also be understood in terms
of the spatial curvature \cite{Rasanen:2008a}.)
For this reason, calling the effect of backreaction a
change of background as in
\cite{Kolb:2009, Clarkson:2009a, Clarkson:2009b}
is misleading. A FRW model which reproduces the
average expansion rate of a clumpy model is better called
a ``fitting model'' or something similar. The metric associated
with it cannot be used to calculate quantities other than
the one specifically fitted for, and usual perturbation theory
around it does not make sense.

In particular, if backreaction is important,
the relation between the expansion rate
and the angular diameter distance given by \re{DAbareq}
(assuming that discreteness is not important) is
different from the FRW case \cite{Rasanen:2008b}\footnote{In contrast,
the relation $D_L=(1+z)^2 D_A$ for the luminosity distance $D_L$
is universal \cite{Ellis:1971}, \cite{Schneider:1992} (page 111),
\cite{Etherington:1933}.}.
Even though either the change of the expansion rate or the change
of the distance due to backreaction can be reproduced in a FRW
model by introducing extra sources of matter or changing the
Einstein equation, it is not possible to do both
at the same time, since FRW models cannot mimic
the correlation \re{DAbareq}.
In a clumpy space, if the dust approximation holds, the distance
is uniquely determined by the function $H(\bz)$ and
the matter density today \cite{Rasanen:2008b}.
(Note that fitting the distance observations may not
necessarily require accelerating expansion, because the relation
between $H$ and $D_A$ is different from the FRW case.)
Analogously, in a FRW universe with general matter content,
the distance is determined by $H(z)$ and the spatial curvature
today \cite{Clarkson:2007b}.
In LTB models, the relationship is different from
either of these cases \cite{February:2009}.

This prediction for both FRW models and backreaction
can be tested with independent observations of distance
and the expansion rate \cite{Avgoustidis:2009}.
The deviation of the backreaction case from the FRW
consistency condition is related to the
difference of the average expansion rate from the FRW
case with vacuum energy and dust \cite{Rasanen:2008b}.
There are relatively good constraints on the distance scale as
a function of redshift from type Ia supernova observations
\cite{FRWexp}, but measurements of the expansion rate
using the ages of passively evolving galaxies are less
precise \cite{ages}. The expansion rate at different
redshifts also enters into the radial mode of the baryon
acoustic oscillations, and a measurement
was reported in \cite{BAO2} (see also
\cite{SylosLabini:2009, MiraldaEscude:2009}).
With better observations of the expansion rate, it
will be possible to more tightly test the statement that
the universe is described by a FRW metric,
independent of the possible existence of exotic matter
or modified gravity \cite{Clarkson:2007b, Shafieloo:2009}\footnote{Note
that this is not a test of the Copernican principle.
The statement that our position in the universe is
not untypical is different from the statement that the
metric is FRW. In fact, the Copernican principle says
nothing about the metric.}.
Similarly, backreaction can be tested without having a
prediction for the average expansion rate.
In the case of backreaction, it helps that there are
independent observational constraints on the matter density today,
while the only way to determine the spatial curvature of a
FRW universe is to make independent measurements
of the expansion rate and distance, and use the FRW relation
between them. (In particular, the CMB has no model-independent
sensitivity to the spatial curvature.)

\subsection{Conclusions} \label{sec:disc}

\para{Summary.}

In \cite{Rasanen:2008b} it was argued that light
propagation can be approximately described in terms
of the overall geometry (meaning the average expansion rate
and other average scalar quantities) in statistically homogeneous
and isotropic dust universes where the distribution
evolves slowly compared to the time it takes for
light to cross the homogeneity scale.
The calculation was incomplete because it
was simply assumed that the light deflection is small,
and there was no result for the amplitude of the image shear.
It was also assumed that the matter is dust and irrotational,
while it is known that such a description does not locally
hold everywhere.

We have now considered general matter content,
with a general observer velocity and a general
hypersurface of statistical homogeneity and isotropy,
with a slowly evolving distribution.
From these assumptions about the spacetime symmetry,
we find that the propagation of typical light rays
over distances longer than the homogeneity scale
can to leading order be treated in terms of a
few average quantities, as in the irrotational dust case.
The redshift is given by the average volume
expansion rate of the hypersurface of statistical
homogeneity and isotropy, assuming that the velocity difference
between the frame of statistical homogeneity and
isotropy and the observer frame is non-relativistic.
The relevant averaging hypersurface is selected by the
symmetry of the situation as the one of statistical
homogeneity and isotropy, while the relevant
velocity is that of the observers, because it
gives the observed redshift.
The angular diameter distance is to leading order determined
by the averages of the expansion rate, time dilation,
energy density and pressure.
The light deflection and the image shear are small.

We have also written down the generalisation of the
Buchert equations \cite{Buchert:1999} for the
evolution of the average expansion rate to general
matter content and averaging hypersurface.
Provided that the difference between frame
of statistical homogeneity and isotropy and the
observer frame is small, and that pressure,
energy flux and anisotropic stress are not significant
in a large fraction of space, we recover the Buchert
equations for dust.

\para{Outlook.}

If backreaction has a large effect on the average
expansion rate, the relation between
the expansion rate and the angular diameter distance
is different from the FRW case, and the difference
grows with the deviation of the expansion rate
from that of the FRW model with dust and vacuum energy.
This is a distinct prediction of backreaction, which
cannot be mimicked by any FRW model \cite{Rasanen:2008b}.

Our arguments are qualitative, and 
should be followed up by a more rigorous quantitative study.
In particular, evaluating the deviations from the mean is
necessary to study lensing and the low multipoles of the CMB.
The equations we have written in the covariant formalism
contain all general relativistic effects and are exact, except
for the geometrical optics approximation.
The study of light propagation in a general
spacetime is reduced to the system of coupled ordinary differential
equations \re{Eder}, \re{eeq}, \re{patsigma2} and \re{Raynull}.
To obtain a solution, we do not have to know
the global geometry, it is only necessary to specify
the distribution along the null geodesic. This approach was
used in the perturbative case in \cite{Kainulainen:2009}.
This formulation is well-suited to slowly evolving statistically
homogeneous and isotropic universes, where the solution is
expected to depend only on the statistical properties of the
distribution. In particular, because in the real universe the
distribution originates from small almost Gaussian fluctuations, its
statistics are even at late times determined by the
initial power spectrum, processed by gravity.
The effect of the discreteness of the matter content of the
universe on light propagation should be clarified.
Finally, even though the relation between the expansion rate
and the distance is already a prediction which can be checked,
it remains of central importance to derive the average expansion
rate from the statistics of structure formation \cite{Rasanen:2008a}
in a reliable manner to allow easier and more comprehensive
comparison with observations.

\acknowledgments

I thank Thomas Buchert, Timothy Clifton, Ruth Durrer, Pedro
Ferreira and David Wiltshire for discussions and disagreements, 
and Thomas Buchert also for comments about the manuscript,
and the Helsinki Institute of Physics and the Tuorla Observatory
for hospitality.

This paper is dedicated to Heidi Hopeamets\"{a}.\\

\end{document}